\documentclass[12pt,draftcls,onecolumn]{IEEEtran}
\usepackage{graphicx,subfigure,amsmath,amssymb}

\begin{document}

\title{Flocking of Multi-agent Dynamical Systems Based on Pseudo-leader Mechanism }

\author{Jin Zhou,~
        Wenwu Yu,~
        Xiaoqun Wu,~
        Michael Small,~
        Jun-an~Lu
\thanks{
        This work was supported by the funding of a Competitive Earmarked Research Grant (No. PolyU 5269/06E)
        from the University Grants Council of Hong Kong. Ms Zhou Jin is supported by the National
        Natural Science Foundation of China under Grant No. 70771084,
        60574045, and the National Basic Research
        Program of China under Grant No. 2007CB310805.}
\thanks{J. Zhou is with the School of Mathematics and Statistics, Wuhan
University, Wuhan 430072, China. J. Zhou was also with the
Department of Electronic and Information Engineering, Hong Kong
Polytechnic University, Hong Kong.}
\thanks{M. Small is with the Department
of Electronic and Information Engineering, Hong Kong Polytechnic
University, Hong Kong.}
\thanks{W. W. Yu is with the Department of Electronic Engineering, City University of Hong
Kong, Hong Kong.}
\thanks{ X. Q. Wu and J. A. Lu are with the School of Mathematics and Statistics, Wuhan University,
Wuhan 430072, China.}
\thanks{(Corresponding author e-mail: enjzhou@gmail.com).}}


\maketitle

\begin{abstract}

Flocking behavior of multiple agents can be widely observed in
nature such as schooling fish and flocking birds. Recent literature
has proposed the possibility that flocking is possible even only a
small fraction of agents are informed of the desired position and
velocity. However, it is still a challenging problem to determine
which agents should be informed or have the ability to detect the
desired information. This paper aims to address this problem. By
combining the ideas of virtual force and pseudo-leader mechanism,
where a pseudo-leader represents an agent who can detect the desired
information, we propose a scheme for choosing pseudo-leaders in a
multi-agent group. The presented scheme can be applied to a
multi-agent group even with an unconnected or switching neighbor
graph. Experiments are given to show that the methods presented in
this paper are of high accuracy and perform well.

\end{abstract}


\IEEEpeerreviewmaketitle

\section{Introduction}

Flocking is a collective behavior of a large group of mobile
agents. Typical flocking phenomena include flocks of birds,
schools of fish, herds of animals and colonies of bacteria. In
nature, animals achieve flocking for various reasons. For example,
in order to keep safety in numbers and also to confuse predators,
they will form flocks for protection. An agent is more likely to
be attacked if it strays away from the flocking group.

As a common demonstration of emergence and emergent behavior,
flocking was first simulated on a computer by Craig Reynolds
\cite{Reynolds}. In 1987, he started with a boid model to build a
simulated flock and introduced three rules to simulate flocking:

\begin{itemize}

\item Collision Avoidance: steer to avoid collision with
nearby flockmates (short range repulsion).

\item Velocity Matching:
steer to match velocity with nearby flockmates.

\item Flock Centering: steer to stay close to nearby flockmates (long
range attraction).

\end{itemize}

\noindent The mechanism, known as ``separation", ``alignment" and
``cohesion", results in all agents moving in a formation with the
same heading and a fixed network structure. From then on, these
three rules have been widely used to study flocking behavior.

As a special case of Reynolds' model, in \cite{Vicsek}, Vicsek
\emph{et al.} presented a simulation model based on nearest
neighborhood law, in which each agent's heading is updated by the
average of the headings of its nearest neighbors and itself. It was
shown that the headings of all the group agents converge to a common
value.

A lot of works have been published based on Reynolds' and Vicsek's
models in recent years \cite{Shen2008}-\cite{XHLi}. To thoroughly
and systematically investigate flocking behavior, artificial
potential functions (APFs) are widely used. In
\cite{Tanner2003_1}, Tanner \emph{et al.} presented an APF in a
network with fixed topology which is a differential, nonnegative
and radially unbounded function of the distance between two
agents. Then, he modified the APF to a nonsmooth one in a network
with switching topology which captures the fact that there is no
agent interaction beyond a proper distance \cite{Tanner2003_2}.
Later, Olfati-Saber modified Tanner's APF and defined another APF
which is a bounded and smooth one for switching topology.

In view of the pitfall of regular fragmentation \cite{OS2006},
which is a phenomenon of flocking failure most likely occurring
for generic set of initial states and large number of agents,
Olfati-Saber also introduced a flocking mechanism based on a
virtual leader \cite{OS2006}. Even though the initial states is
selected randomly, the mechanism can guarantee a flocking
behavior. However, this method assumes that all the agents know
information about the virtual leader. In consideration of
practical use, Shi \emph{et al.} and Su \emph{et al.} showed that
flocking appears even when only some agents are informed
\cite{HSS_1}-\cite{HS}, these we call pseudo-leaders in this
paper.

Though Shi \emph{et al.} and Su \emph{et al.} showed that flocking
could appear when not all the agents are informed in a group,
precisely which agents should be informed has not yet been
considered. In this paper, we focus on investigating which agents
should be selected as pseudo-leaders for flocking in a weighted
network. Another difference from previous work is that, in the
absence of control, the acceleration of each agent is dynamic.
This can be seen in our system model in Section II. That is, if
there are no attractive/repulsive control, no information exchange
with others and no information received from the virtual leader
for an agent, its acceleration is dynamic instead of constant, as
is the case in previous work. By using Lyapunov stability theory,
a simple criterion for choosing pseudo-leaders is proposed.

The reminder of this paper is organized as follows. In Section II,
model depiction, preliminaries about graph theory, and
mathematical analysis are briefly introduced. In Section III, the
main results are proposed. We investigate how to select
pseudo-leaders for flocking in a multi-agent dynamical group with
fixed topology. Section IV gives some extensions and discussion of
the main results so that one can gain useful insight into the
problem of choosing pseudo-leaders. Two computational examples in
small-sized and large-scale groups are simulated to illustrate
effectiveness of the proposed approach in Section V. We summarize
the main ideas and conclusions in Section VI.

\section{Preliminaries}

\subsection{Graph theory}

To make this paper self-contained, some basics of graph theory are
recalled \cite{Godsil}.

A \emph{graph} $\mathcal{G}$ is a pair of sets
$(\mathcal{V},\mathcal{E})$, where $\mathcal{V}$ is a finite
non-empty set of elements called \emph{vertices}, and
$\mathcal{E}$ is a set of unordered pairs of distinct vertices
called \emph{edges}. The set $\mathcal{V}$ and $\mathcal{E}$ are
the vertex set and edge set of $\mathcal{G}$, and are often
denoted by $\mathcal{V(G)}$ and $\mathcal{E(G)}$, respectively. If
$i,j\,\in\,\mathcal{V}$ and $(i,j)\,\in\,\mathcal{E}$, then $i$
and $j$ are \emph{adjacent vertices}, or \emph{neighbors}.
\emph{The set of neighbors of a vertex} $i$ is $\mathcal{N}_i$. A
\emph{walk} in a graph is a sequence of vertices and edges $j_0$,
$e_1$, $j_1$, $\cdots$, $e_k$, $j_k$, in which each edge
$e_r\,=\,(j_{r-1},j_r)$. A \emph{path} is a walk in which no
vertex is repeated. If there is a path between any two vertices of
a graph $\mathcal{G}$, then $\mathcal{G}$ is said to be
\emph{connected}. If $\mathcal{G}$ and $\mathcal{G}_1$ are graphs
with $\mathcal{V}(\mathcal{G}_1)\,\subseteq\,\mathcal{V(G)}$ and
$\mathcal{E}(\mathcal{G}_1)\,\subseteq\,\mathcal{E(G)}$, then
$\mathcal{G}_1$ is a \emph{subgraph} of $\mathcal{G}$.

The position and velocity neighbor graph
$\mathcal{G}\,=\,(\mathcal{V},\mathcal{E},\textbf{W})$ is a
weighted graph consisting of a set of indexed vertices and a set
of ordered edges, where
$\textbf{W}\,=\,(w_{ij})\,\in\,\mathbb{R}^{N\,\times\,N}$ is the
weight matrix which represents the weighted coupling coefficients
of interaction between the agents. If there is a link from vertex
$i$ to vertex $j\,(j\,\neq\,i)$, then $w_{ij}\,=\,w_{ji}\,>\,0$
and $w_{ij}$ is the weight; otherwise, $w_{ij}\,=\,0$. Throughout
the paper, assume that $\textbf{W}$ is a symmetric matrix
satisfying \emph{diffusive condition}

$$
\sum\limits_{j=1}^N\,w_{ij}\,=\,0.
$$

\subsection{Model depiction}

Consider a multi-agent system consisting of $N$ agents. Here, the
moving model of each agent in the group is given by

\begin{equation}\label{1}
\begin{array}{l}
    \dot{\textbf{p}}_i\,=\,\textbf{v}_i,\\
    \dot{\textbf{v}}_i\,=\,\textbf{f}(\textbf{v}_i)\,+\,\textbf{u}_i,
\end{array}
\end{equation}

\noindent where $1\leq{i}\leq{N}$,
$\textbf{p}_i\,\in\,\mathbb{R}^n$,
$\textbf{v}_i\,\in\,\mathbb{R}^n$ and
$\textbf{f}(\textbf{v}_i)\,\in\,\mathbb{R}^n$ denote the position,
velocity and the acceleration dynamics (without control input) of
the $i$-th agent respectively, and
$\textbf{u}_i\,\in\,\mathbb{R}^n$ is the control input of agent
$i$.

The motion model for the virtual leader is

\begin{equation}\label{2}
\begin{array}{l}
    \dot{\textbf{p}}_l\,=\,\textbf{v}_l,\\
    \dot{\textbf{v}}_l\,=\,\textbf{f}(\textbf{v}_l),
\end{array}
\end{equation}

\noindent where $\textbf{p}_l\,\in\,\mathbb{R}^n$,
$\textbf{v}_l\,\in\,\mathbb{R}^n$ and
$\textbf{f}(\textbf{v}_l)\,\in\,\mathbb{R}^n$ represent the
position, velocity and the acceleration dynamics of the virtual
leader, respectively.

The flocking task is to design an appropriate control input
$\textbf{u}_i$ such that Reynolds' rules are followed, and then
all agents moving in a formation with a common heading and
collision avoidance.

\subsection{Mathematical Preliminaries}

In this subsection, some useful mathematical definitions, lemmas
and assumptions are outlined.

\vspace*{5mm} \noindent \textbf{Definition 1.} A matrix
$\textbf{A}\,=\,(a_{ij})\,\in\,\mathbb{R}^{N\,\times\,N}$ is
called \emph{reducible} if the indices $1,2,\cdots,N$ can be
divided into two disjoint nonempty sets
$i_1,i_2,\cdots,i_{\beta_1}$ and $j_1,j_2,\cdots,j_{\beta_2}$
(with $\beta_1\,+\,\beta_2\,=\,N$) such that

$$
a_{i_{\kappa_1},j_{\kappa_2}}\,=\,0
$$

\noindent for $1\,\leq\,\kappa_1\,\leq\,\beta_1$ and
$1\,\leq\,\kappa_2\,\leq\,\beta_2$. A matrix
$\textbf{A}\,=\,(a_{ij})\,\in\,\mathbb{R}^{N\,\times\,N}$ is
\emph{irreducible} if and only if it is not reducible. \vspace*{5mm}

A matrix is reducible if and only if it can be placed into block
upper-triangular form by simultaneous row/column permutations. In
addition, a matrix is reducible if and only if its associated
graph is not connected \cite{Gradshteyn}.

\vspace*{5mm} \noindent \textbf{Definition 2} \cite{Horn}
\textbf{.} A matrix
$\textbf{A}\,=\,(a_{ij})\,\in\,\mathbb{R}^{N\,\times\,N}$ is
called \emph{diagonally dominant} if
$|a_{ii}|\,\geq\,\,\sum\limits_{j=1,j\,\neq\,i}^N\,|a_{ij}|$ for
$1\,\leq\,i\,\leq\,N$. $\textbf{A}$ is called \emph{strictly
diagonally dominant} if
$|a_{ii}|\,>\,\,\sum\limits_{j=1,j\,\neq\,i}^N\,|a_{ij}|$ for
$1\,\leq\,i\,\leq\,N$. \vspace*{5mm}

The Gershgorin circle theorem  \cite{Horn} results in many
interesting conclusions. A strictly diagonally dominant matrix is
nonsingular. A symmetric diagonally dominant real matrix with
nonnegative diagonal entries is positive semi-definite. If a
symmetric matrix is strictly diagonally dominant and all its
diagonal elements are positive, then its eigenvalues are positive;
if all its diagonal elements are negative, then its eigenvalues
are negative. Thus it is obvious that the real part of eigenvalues
of the weight matrix $\textbf{W}$ which are all negative except an
eigenvalue $0$ with multiplicity one. Throughout the paper, we
denote a positive definite (positive semi-definite, negative
definite, negative semi-definite) matrix $\textbf{A}$ as
$\textbf{A}>\textbf{0}\;(\geq\textbf{0},\;<\textbf{0},\;\leq\textbf{0})$.

\vspace*{5mm} \noindent \textbf{Definition 3} \cite{Hassan2002}
\textbf{.} A set $\Omega$ is said to be a \emph{positively
invariant set} with respect to an equation if a solution $x(t)$ of
this equation satisfies

$$
x(0)\,\in\,\Omega\,\Rightarrow\,x(t)\,\in\,\Omega\ \ \ \ \ \ \ \ \
\ \forall\,t\,\geq\,0.
$$

\vspace*{5mm} \noindent \textbf{Definition 4} \cite{Hassan2002}
\textbf{.} A continuous function
$g:[0,\bar{\rho})\rightarrow[0,\infty)$ is said to belong to
\emph{class $\mathcal{K}$} if it is strictly increasing and
$g(0)=0$. It is said to belong to \emph{class
$\mathcal{K}_\infty$} if $\bar{\rho}=\infty$ and
$g(\rho)\rightarrow\infty$ as $\rho\rightarrow\infty$.

\vspace*{5mm} \noindent \textbf{Definition 5.} \emph{$2$-norm},
also called \emph{Euclidean norm}, of a vector $\xi$ is defined as

$$
\|\xi\|_2\,=\,\sqrt{\xi^{\top}\,\xi},
$$

\noindent where $\top$ denotes the transpose of a vector or a
matrix.

\vspace*{5mm} \noindent \textbf{Definition 6} \cite{OS2006}
\textbf{.} \emph{$\sigma$-norm} of a vector $\xi$ is defined as

$$
\|\xi\|_\sigma\,=\,\frac{1}{\sigma}\,\left[\sqrt{1\,+\,\sigma\,\|\xi\|_2^2}-1\right],
$$

\noindent where $\sigma$ is a positive constant. \vspace*{5mm}

Due to the fact that $\|\xi\|_\sigma=0$ as $\|\xi\|_2=0$,
$\|\xi\|_\sigma\rightarrow\infty$ as $\|\xi\|_2\rightarrow\infty$,
and that
$\frac{d\,\|\xi\|_\sigma}{d\,\|\xi\|_2}\,=\,\frac{\|\xi\|_2}{\sqrt{1\,+\,\sigma\,\|\xi\|_2^2}}\,>\,0$
when $\|\xi\|_2\,\neq\,0$, $\|\xi\|_\sigma$ is a class
$\mathcal{K}_\infty$ function of $\|\xi\|_2$. On the other hand,
note that even though $\|\xi\|_\sigma$ is not a norm in the sense of
algebra, it is differentiable everywhere while $\|\xi\|_2$ is not.
Thus $\|\xi\|_\sigma$ instead of $\|\xi\|_2$ is used to construct
APF in this paper.

\vspace*{5mm} \noindent \textbf{Definition 7.} \emph{Artificial
Potential Function (APF)} $V_{ij}$ is a differentiable,
nonnegative, radially unbounded \cite{Hassan2002} function of
$\|\textbf{p}_i-\textbf{p}_j\|_\sigma$, which is the $\sigma$-norm
of the position error between the $i$-th and the $j$-th agents.
$V_{ij}(\|\textbf{p}_i-\textbf{p}_j\|_\sigma)$ has the following
properties:

\noindent (1)\;
$V_{ij}(\|\textbf{p}_i-\textbf{p}_j\|_\sigma)\,\rightarrow\,\infty$
as $\|\textbf{p}_i-\textbf{p}_j\|_\sigma\,\rightarrow\,0$,

\noindent (2)\; $V_{ij}(\|\textbf{p}_i-\textbf{p}_j\|_\sigma)$
attains its unique minimum when the $i$-th agent and the $j$-th
agent are located at a desired distance. \vspace*{5mm}

An example of APF is

\begin{equation}\label{3}
V_{ij}(\|\textbf{p}_i-\textbf{p}_j\|_\sigma)\,=\,c_1\,\ln\,\|\textbf{p}_i-\textbf{p}_j\|_\sigma^2\,
+\,\frac{c_2}{\|\textbf{p}_i-\textbf{p}_j\|_\sigma^2},
\end{equation}

\noindent where $c_1>0$, $c_2>0$ are two constants. The potential
function approaches infinity as
$\|\textbf{p}_i-\textbf{p}_j\|_\sigma$ tends to $0$, and attains
its unique minimum when
$\|\textbf{p}_i-\textbf{p}_j\|_\sigma\,=\,\sqrt{\frac{c_2}{c_1}}$.

In order to derive the main results, the following lemmas and
assumption are needed.

\vspace*{5mm} \noindent \textbf{Lemma 1.} (Schur complement
\cite{Boyd,WWYu2008}) The following linear matrix inequality (LMI)

\vspace{-5mm}
$$
    \left(\begin{array}{ll}
      \mathcal{A}(x) & \mathcal{B}(x)\\
      \mathcal{B}(x)^{\top} & \mathcal{C}(x)
    \end{array}\right)<\textbf{0},
$$

\noindent where $\mathcal{A}(x)^{\top}\,=\,\mathcal{A}(x)$,
$\mathcal{C}(x)^{\top}\,=\,\mathcal{C}(x)$, is equivalent to
either of the following conditions:

\noindent (a) $\mathcal{A}(x)<\textbf{0}$ and
$\mathcal{C}(x)\,-\,\mathcal{B}(x)^{\top}\,\mathcal{A}(x)^{-1}\,
\mathcal{B}(x)<\textbf{0}$\\
\noindent (b) $\mathcal{C}(x)<\textbf{0}$ and
$\mathcal{A}(x)\,-\,\mathcal{B}(x)\,\mathcal{C}(x)^{-1}\,\mathcal{B}(x)^{\top}<\textbf{0}$.

\vspace*{5mm} \noindent \textbf{Lemma 2} \cite{WLLu} \textbf{.} If a
symmetric matrix
$\textbf{A}\,=\,(a_{ij})\,\in\,\mathbb{R}^{N\,\times\,N}$
$(a_{ij}>0,\;{i}\neq{j})$ is irreducible and satisfies diffusive
condition, then $\textbf{A}-\textbf{D}_i<\textbf{0}$ holds for a
diagonal matrix
$\textbf{D}_i\,=\,diag\{0,\cdots,0,\delta,0,\cdots,0\}\,\in\,\mathbb{R}^{N\,\times\,N}$,
where the $i$-th $(1\,\leq\,i\,\leq\,N)$ element $\delta$ is any
positive constant and the others are $0$.

\vspace*{5mm} \noindent \textbf{Assumption 1 (A1).} Suppose that
there exists a positive constant $\alpha$ satisfying
$(\xi_2-\xi_1)^{\top}(\textbf{f}(\xi_2)-\textbf{f}(\xi_1))\,\leq\,\alpha\,\|\xi_2-\xi_1\|_2^2$
for any two vectors $\xi_1,\xi_2\,\in\,\mathbb{R}^n$.
\vspace*{5mm}

\section{Main results}

In this section, a group of mobile agents with fixed topology is
considered. Assume that the position and velocity neighbor graph
$\mathcal{G}\,=\,(\mathcal{V},\mathcal{E},\textbf{W})$ is
connected. That is, the weight matrix $\textbf{W}$ is irreducible.
For the case that
$\mathcal{G}\,=\,(\mathcal{V},\mathcal{E},\textbf{W})$ is
unconnected, discussion will be found in Section IV. C.

Generally, there should be attractive and repulsive mechanisms in
the control input, where the attraction indicates that each agent
wants to be close to nearby agents and repulsion provides the fact
that each agent does not want to be too close to nearby
flockmates. These two mechanisms can be jointly embodied in APF
(actually there are many ways to achieve attraction and
repulsion).

Besides ``separation" and ``cohesion",  ``alignment" is an
important rule in flocking. We use the approach that group members
receive moving information from a virtual leader to realize
alignment. Based on this approach, recent literatures have shown
the phenomenon that flocking will appear even if not all the
agents are informed in the group. Here, we investigate which
agents should be selected as informed ones for flocking in a
weighted network. An agent, which utilizes the motion information
of the virtual leader as a reference in its controller, is a
pseudo-leader of the group. Suppose that the $i_1$-th, $i_2$-th,
$\cdots,i_m$-th agents are chosen as the pseudo-leaders.

The control law for the $i$-th agent is given by

\begin{equation}\label{4}
\begin{array}{l}
    \textbf{u}_i\,=\,\textbf{u}^1_i\,+\,\textbf{u}^2_i\,+\,\textbf{u}^3_i,\\
    \textbf{u}^1_i\,=\,-\sum\limits_{j\,\in\,\mathcal{N}_i}\,w_{ij}(\textbf{v}_i-\textbf{v}_j)\,
    =\,\sum\limits_{j=1}^N\,w_{ij}\textbf{v}_j,\\
    \textbf{u}^2_i\,=\,-\sum\limits_{j\,\in\,\mathcal{N}_i}\,
    w_{ij}\,\nabla_{\textbf{p}_i}V_{ij}(\|\textbf{p}_i-\textbf{p}_j\|_\sigma)\,
    =\,-\sum\limits_{j=1,j\neq{i}}^N\,
    w_{ij}\,\nabla_{\textbf{p}_i}V_{ij}(\|\textbf{p}_i-\textbf{p}_j\|_\sigma),\\
    \textbf{u}^3_i\,=\,
    \left\{ \begin{array}{l}
    -h_{i_r}(\textbf{v}_{i_r}-\textbf{v}_l),\ \
    \dot{h}_{{i_r}}\,=\,k_{{i_r}}(\|\textbf{v}_{i_r}-\textbf{v}_l\|_2^2)\ \ \ \
    \ i\,\in\,\{i_r|1\leq{r}\leq{m},\;1\leq{i_r}\leq{N}\},\\
    \textbf{0}\ \ \ \ \ \ \ \ \ \ \ \ \ \ \ \ \ \ \ \ \ \ \ \ \ \
    \ \ \ \ \ \ \ \ \ \ \ \ \ \ \ \ \ \ \ \ \ \ \ \
    \mathrm{otherwise},
    \end{array}\right.
\end{array}
\end{equation}


\noindent where $\nabla$ represents the gradient of a function,
$h_{i_r}$ is the adaptive feedback gain,
$k_{i_r}\,>\,0\;(1\,\leq\,r\,\leq\,m,1\;\leq\,i_r\,\leq\,N)$ is a
constant. For agent $i$, $\textbf{u}^1_i$ is the velocity coupling
of interactions between agents, and $\textbf{u}^2_i$ stands for
the coupling gradient of APF with respect to its position. The
term $\textbf{u}^3_i$, which will disappear if agent $i$ is not a
pseudo-leader, controls the received information from the virtual
leader. It provides an adaptive feedback adjusting mechanism and
will be more practical in engineering than those with linear
feedback ones.

Define the position and velocity error between agent $i$ and the
virtual leader as
$\textbf{e}_i^\textbf{p}\,\triangleq\,\textbf{p}_i\,-\textbf{p}_l$,
$\textbf{e}_i^\textbf{v}\,\triangleq\,\textbf{v}_i\,-\textbf{v}_l$,
then we have

$$
\begin{array}{l}
\dot{\textbf{e}}_i^\textbf{p}\,=\,\textbf{e}_i^\textbf{v},\\
\dot{\textbf{e}}_i^\textbf{v}\,=\,\textbf{f}(\textbf{v}_i)\,-\,\textbf{f}(\textbf{v}_l)\,+\,\textbf{u}_i.
\end{array}
$$

\noindent Since A1 holds, we obtain

$$
{\textbf{e}_i^\textbf{v}}^{\top}(\textbf{f}(\textbf{v}_i)-\textbf{f}(\textbf{v}_l))\,
\leq\,\alpha\,\|\textbf{e}_i^\textbf{v}\|_2^2
$$

\noindent for $1\,\leq\,i\,\leq\,N$. In addition, it is easily to
get

$$
\begin{array}{rcl}
\nabla_{\textbf{p}_i}V_{ij}(\|\textbf{p}_i-\textbf{p}_j\|_\sigma)\,
& = & \,\nabla_{\textbf{p}_i}\,V_{ij}(\|\textbf{e}^\textbf{p}_i-\textbf{e}^\textbf{p}_j\|_\sigma)\\
& = &
\,\nabla_{\textbf{e}_i^\textbf{p}}V_{ij}(\|\textbf{e}^\textbf{p}_i-\textbf{e}^\textbf{p}_j\|_\sigma)
\end{array}
$$

\noindent and

$$
\nabla_{\textbf{e}_i^\textbf{p}}V_{ij}(\|\textbf{e}^\textbf{p}_i-\textbf{e}^\textbf{p}_j\|_\sigma)\,
=\,-\nabla_{\textbf{e}_j^\textbf{p}}V_{ij}(\|\textbf{e}^\textbf{p}_i-\textbf{e}^\textbf{p}_j\|_\sigma),
$$

\noindent and then it follows

$$
\begin{array}{rcl}
\frac{d}{dt}\,V_{ij}(\|\textbf{p}_i-\textbf{p}_j\|_\sigma)\,
& = &
\,2\,\dot{\textbf{e}}_i^{\textbf{p}\,\top}\,
\nabla_{\textbf{e}_i^\textbf{p}}\,V_{ij}(\|\textbf{e}^\textbf{p}_i-\textbf{e}^\textbf{p}_j\|_\sigma)\\
& = &
\,2\,{\textbf{e}_i^\textbf{v}}^\top\,\nabla_{\textbf{p}_i}\,V_{ij}(\|\textbf{p}_i-\textbf{p}_j\|_\sigma).
\end{array}
$$

\vspace*{5mm} \noindent \textbf{Theorem 1.} Suppose that A1 holds.
If $\textbf{W}_{N-m}\,+\,\alpha\,\textbf{I}_{N-m}\,<\,\textbf{0}$,
where $\textbf{W}_{N-m}$ is the minor matrix of the weight matrix
$\textbf{W}$ by removing all the $i_r$-th
$(1\,\leq\,r\,\leq\,m,1\;\leq\,i_r\,\leq\,N)$ row-column pairs,
flocking behavior appears in system (1) by the control strategy
(2) and (4). That is, the velocities of all agents approach the
desired velocity asymptotically, collisions between agents are
avoided and the distances between all agents are invariant.
Furthermore, the global potentials of all agents in the group are
minimized with the final configuration.

\vspace*{5mm} \noindent \emph{Proof.} Consider a positive
semi-definite function as

$$
\begin{array}{rcl}
L & = &
\frac{1}{2}\,\sum\limits_{i=1}^N\,\sum\limits_{j=1,j\neq{i}}^N\,
w_{ij}\,V_{ij}(\|\textbf{p}_i-\textbf{p}_j\|_\sigma)\,
+\,\frac{1}{2}\,\sum\limits_{i=1}^N\,{\textbf{e}_i^\textbf{v}}^\top\,\textbf{e}_i^\textbf{v}\\
&   &
+\,\frac{1}{2}\,\sum\limits_{r=1}^m\,\frac{(h_{i_r}\,-\,h)^2}{k_{i_r}},
\end{array}
$$

\noindent where $h>0$ is a constant to be determined.

We now prove that
$\Omega_c\,=\,\{(\textbf{p}_i-\textbf{p}_j,\textbf{e}_i^\textbf{v})\,|\,L\,\leq\,c,\;c>0\}$,
the sub-level set of $L$, is compact. Firstly, from $L\,\leq\,c$
one gets $\|\textbf{e}_i^\textbf{v}\|_2^2\,
=\,{\textbf{e}_i^\textbf{v}}^\top\,\textbf{e}_i^\textbf{v}\,\leq\,2\,c$
and
$w_{ij}\,V_{ij}(\|\textbf{p}_i-\textbf{p}_j\|_\sigma)\,\leq\,2\,c$
for $1\,\leq\,i,j\,\leq\,N$. Due to the fact that the continuous
function $V_{ij}$ is radially unbounded,
$V_{ij}(\|\textbf{p}_i-\textbf{p}_j\|_\sigma)\,\rightarrow\,\infty$
as $\|\textbf{p}_i-\textbf{p}_j\|_\sigma\,\rightarrow\,0$, and
that $\|\textbf{p}_i-\textbf{p}_j\|_\sigma$ is a class
$\mathcal{K}_\infty$ function with respect to
$\|\textbf{p}_i-\textbf{p}_j\|_2$, there exists a positive
constant $c'$ such that
$\|\textbf{p}_i-\textbf{p}_j\|_2\,\leq\,c'$. Thus $\Omega_c$ is a
bounded set. Secondly, because
$\Omega'_c\,=\,\{L\,|\,L\,\leq\,c,\;c>0\}$ is closed, $\Omega_c$
is a closed set for the continuity of function $L$. Then according
to Heine-Borel Theorem \cite{Jeffreys}, $\Omega_c$ is compact.

Regarding $L$ as a Lyapunov candidate, its derivative along the
trajectories of (1), (2) and (4) is

$$
\begin{array}{rcl}
\dot{L} & = &
    \sum\limits_{i=1}^N\,{\textbf{e}_i^\textbf{v}}^\top\,
    \left(\,\textbf{f}(\textbf{v}_i)-\textbf{f}(\textbf{v}_l)\,
    +\,\sum\limits_{j=1}^N\,w_{ij}\,\textbf{e}_j^\textbf{v}\,-\sum\limits_{j=1,j\neq{i}}^N\,
    w_{ij}\,\nabla_{\textbf{p}_i}V_{ij}(\|\textbf{p}_i-\textbf{p}_j\|_\sigma)\right)\\
 &   & +\,\sum\limits_{i=1}^N\sum\limits_{j=1,j\neq{i}}^N\,
    w_{ij}\,{\textbf{e}_i^\textbf{v}}^\top\,\nabla_{\textbf{p}_i}V_{ij}(\|\textbf{p}_i-\textbf{p}_j\|_\sigma)\,
    -\,\sum\limits_{r=1}^m\,h_{i_r}\,{\textbf{e}_{i_r}^\textbf{v}}^\top\,\textbf{e}_{i_r}^\textbf{v}\,
    +\,\sum\limits_{r=1}^m\,(h_{i_r}\,-\,h)\,{\textbf{e}_{i_r}^\textbf{v}}^\top\,\textbf{e}_{i_r}^\textbf{v}\\
 & \leq &
    \sum\limits_{i=1}^N\,\alpha\,{\textbf{e}_i^\textbf{v}}^\top\,\textbf{e}_i^\textbf{v}\,
    +\,\sum\limits_{i=1}^N\,\sum\limits_{j=1}^N\,w_{ij}\,{\textbf{e}_i^\textbf{v}}^\top\,\textbf{e}_j^\textbf{v}\,
    -\,\sum\limits_{r=1}^m\,h\,{\textbf{e}_{i_r}^\textbf{v}}^\top\,\textbf{e}_{i_r}^\textbf{v}\\
 & = &
    \sum\limits_{i=1}^N\,\alpha\,\|\textbf{e}_i^\textbf{v}\|_2^2\,
    +\,\sum\limits_{i=1}^N\,w_{ii}\,\|\textbf{e}_i^\textbf{v}\|_2^2\,
    +\,\sum\limits_{i=1}^N\,\sum\limits_{j=1,j\neq{i}}^N\,w_{ij}\,{\textbf{e}_i^\textbf{v}}^\top\,
    \textbf{e}_j^\textbf{v}\,
    -\,\sum\limits_{r=1}^m\,h\,\|\textbf{e}_{i_r}^\textbf{v}\|_2^2\\
 & \leq &
    \sum\limits_{i=1}^N\,\alpha\,\|\textbf{e}_i^\textbf{v}\|_2^2\,
    +\,\sum\limits_{i=1}^N\,w_{ii}\,\|\textbf{e}_i^\textbf{v}\|_2^2\,
    +\,\sum\limits_{i=1}^N\,\sum\limits_{j=1,j\neq{i}}^N\,w_{ij}\,\|\textbf{e}_i^\textbf{v}\|_2\,
    \|\textbf{e}_j^\textbf{v}\|_2\,
    -\,\sum\limits_{r=1}^m\,h\,\|\textbf{e}_{i_r}^\textbf{v}\|_2^2\\
 & \triangleq &
    {\textbf{e}^\textbf{v}}^\top\,\textbf{Q}\,\textbf{e}^\textbf{v},
\end{array}
$$

\noindent where
$\textbf{e}^\textbf{v}\,=\,(\|\textbf{e}_1^\textbf{v}\|_2,\|\textbf{e}_2^\textbf{v}\|_2,\cdots,
\|\textbf{e}_N^\textbf{v}\|_2)^{\top}$,
$\textbf{Q}\,=\,\alpha\,\textbf{I}_N\,+\,\textbf{W}\,-\,\textbf{H}$,
and $\textbf{H}$ is a diagonal matrix whose $i_r$-th
$(1\leq{r}\leq{m})$ elements are $h$ and the others are $0$.

After applying row-column permutation, $\textbf{Q}$ can be changed
into

$$
\left(
\begin{array}{cc}
\textbf{W}^*+\alpha\textbf{I}_{m}-{h}\textbf{I}_{m} & \mathbf{W}^{**} \\
{\textbf{W}^{**}}^{\top} & \textbf{W}_{N-m}+\alpha\textbf{I}_{N-m}
\end{array}
\right),
$$

\noindent where $\textbf{W}^{*},\textbf{W}^{**}$ are the
corresponding matrices with compatible dimensions. Since
$\textbf{W}_{N-m}\,+\,\alpha\,\textbf{I}_{N-m}\,<\,\textbf{0}$,
$\textbf{W}_{N-m}\,+\,\alpha\,\textbf{I}_{N-m}$ is invertible.
According to Lemma 1, choosing $h$ be a positive constant
satisfying
$\textbf{W}^{*}\,+\,\alpha\,\textbf{I}_{m}\,-\,h\,\textbf{I}_m\,
-\,\textbf{W}^{**}\,(\textbf{W}_{N-m}\,+\,\alpha\,\textbf{I}_{N-m})^{-1}\,{\textbf{W}^{**}}^{\top}\,<\,\textbf{0}$,
we have $\textbf{Q}\,<\,\textbf{0}$. Furthermore,
$\dot{L}\,\leq\,0$, $L(t)$ is a non-increasing function of $t$.
Thus any solution of (1), (2) and (4) starting in $\Omega_c$ will
stay in it. Namely, $\Omega_c$ is a positively invariant set.

Because $\Omega_c$ is compact and positively invariant, every
solution of the system converges to the largest invariant set
$\Omega^*$ of the set
$\{(\textbf{p}_i-\textbf{p}_j,\textbf{e}_i^\textbf{v})\,|\,\dot{L}=0\}$
on the basis of LaSalle's invariance principle \cite{Hassan2002}. In
$\Omega^*$,
$\dot{\textbf{e}}_i^\textbf{p}\,=\,\textbf{e}_i^\textbf{v}\,=\,\textbf{0}$,
which means that all agent velocities are equal and their position
differences remain unchanged in steady state.

Furthermore, $\textbf{e}_i^\textbf{v}\,=\,\textbf{0}$ leads to
$\dot{\textbf{e}}_i^\textbf{v}\,=\,\textbf{0}$ in $\Omega^*$.
Combining with equation (1), (2) and (4), we have
$\sum\limits_{j=1,j\neq{i}}^N\,w_{ij}\,\nabla_{\textbf{p}_i}V_{ij}(\|\textbf{p}_i-\textbf{p}_j\|_\sigma)\,=\,0$
for $1\,\leq\,i\,\leq\,N$. This indicates that the group final
configuration is a local minima of global potential function of
agent $i$.

Collision avoidance can be proved by contradiction. Assume that
there exists a time $t_1>0$ so that the position difference
between two distinct agents ${i}^*$ and ${i}^{**}$ satisfies
$\|\textbf{p}_{{i}^*}-\textbf{p}_{{i}^{**}}\|_2\,\rightarrow\,0$
as $t\,\rightarrow\,t_1$. Then
$\|\textbf{p}_{{i}^{*}}-\textbf{p}_{{i}^{**}}\|_\sigma\,\rightarrow\,0$
since $\|\centerdot\|_\sigma$ is a class $\mathcal{K}_\infty$
function with respect to $\|\centerdot\|_2$. According to the
definition of APF,
$V_{ij}(\|\textbf{p}_{{i}^{*}}-\textbf{p}_{{i}^{**}}\|_\sigma)\,\rightarrow\,\infty$.
This is in contradiction with the fact that $\Omega_c$ is a
positively invariant set. Therefore, no two agents collide at any
time $t\,\geq\,0$.

Thus the proof is completed.   {$\hfill{} \Box$}

\vspace*{5mm} From this theorem, if
$\lambda_{max}(\textbf{W}_{N-m})\,<\,-\alpha$, where
$\lambda_{max}(\centerdot)$ represents the maximum eigenvalue of a
symmetric matrix, $\bar{\mathcal{V}}\,=\,\{i_1,i_2,\cdots,i_m\}$
can be chosen as the pseudo-leader set to guarantee flocking in
system (1), (2) and (4).


\section{Discussion and Extensions}

In this section, some remarks and extensions are discussed to give
some insights into the main results.

\subsection{Other flocking models}

Previous papers studied a classical model for flocking behavior in
such as \cite{Tanner2003_1}-\cite{HSS_2}. The motion models of the
$i$-th agent and the virtual leader respectively are

\begin{equation}\label{5}
\begin{array}{l}
    \dot{\textbf{p}}_i\,=\,\textbf{v}_i,\\
    \dot{\textbf{v}}_i\,=\,\textbf{u}_i,
\end{array}
\end{equation}

\noindent and

\begin{equation}\label{6}
\begin{array}{l}
    \dot{\textbf{p}}_l\,=\,\textbf{v}_l,\\
    \dot{\textbf{v}}_l\,=\,\textbf{0},
\end{array}
\end{equation}

\noindent where $\textbf{p}_i$, $\textbf{v}_i$
$({i}\,=\,1,2,\cdots,{N},l)$,
$\textbf{u}_i({i}\,=\,1,2,\cdots,{N})$ and $\textbf{f}$ have the
same meaning with the equations (1) and (2). It is assumed that if
there are no attractive/repulsive control, no information exchange
with others and no information receiving from the virtual leader
for an agent, then its acceleration is
$\textbf{f}(\textbf{v}_i)\,=\,\textbf{f}(\textbf{v}_l)\,=\,0$. For
the virtual leader, $\textbf{f}(\textbf{v}_l)\,=\,0$ implies that
it moves along a straight line with a desired velocity
$\textbf{v}_l(0)$. Letting $\alpha\,=\,0$, we have the following
theorem for choosing pseudo-leaders in the classical model.

\vspace*{5mm} \noindent \textbf{Theorem 2.} If
$\textbf{W}_{N-m}\,<\,\textbf{0}$ holds, flocking behavior will
appear in system (5) by the control strategy (4) and (6). That is,
the velocities of all agents approach the desired velocity
asymptotically, collisions between agents are avoided and the
distances between all agents are invariant. In addition, the
global potentials of all these agents are minimized with the group
final configuration. \vspace*{5mm}

For other flocking models such as taking velocity damping into
consideration \cite{HS} (which is frequently unavoidable when
objects move with high speeds or in a viscous environment), the
schemes for determining pseudo-leaders can be attained by similar
analysis.


\subsection{A single pseudo-leader is enough for flocking}

Consider the classical model at first. Denote the minor matrix of
$\textbf{W}$ by deleting any row-column pair as
$\textbf{W}_{N-1}$. It can be rewritten as
$\textbf{W}_{N-1}\,=\,\hat{\textbf{W}}_{N-1}\,+\,\sum\limits_i\,\textbf{D}_i$,
where $1\,\leq\,i\,\leq\,\chi$, $1\,\leq\,\chi\,\leq\,N-1$,
$\hat{\textbf{W}}_{N-1}$ is the corresponding symmetric and
diffusive matrix, and $\textbf{D}_i$ is the diagonal matrix where
the $i$-th element is negative and the others are $0$. According
to Lemma 2, $\textbf{W}_{N-1}$ is negative definite. Therefore,
flocking will occur with just one single pseudo-leader (any agent
is available as an option) based on Theorem 2.

Below we will discuss the model presented in Section II. B.
Suppose that the weight matrix $\textbf{W}\,=\,w\,\textbf{B}$,
where $w>0$ is the common weight coupling,
$\textbf{B}\,=\,(b_{ij})\,\in\,\mathbb{R}^{N\,\times\,N}$ is the
adjacent matrix with $b_{ij}\,=\,b_{ji}\,=\,1$ if there is a link
from agent $i$ to agent $j\,(j\,\neq\,i)$ and $b_{ij}\,=\,0$
otherwise. Moreover, $\textbf{B}$ satisfies diffusive condition
$\sum\limits_{j=1}^N\,b_{ij}\,=\,0$. For an agent
$i_r\;(1\,\leq\,r\,\leq\,m,\;1\,\leq\,i_r\,\leq\,N)$, if the minor
matrix $\textbf{B}_{N-1}$ that is obtained by deleting the
$i_r$-th row-column pair of $\textbf{B}$ satisfies
$\lambda_{max}(\textbf{B}_{N-1})\,<\,-\frac{\alpha}{w}\,$, it can
be picked out as a pseudo-leader for flocking according to Theorem
1. Since $\textbf{B}_{N-1}$ is negative definite by Gerschgorin
theorem, it is concluded that flocking will be achieved with just
one single pseudo-leader, which can be selected randomly from the
vertex set $\mathcal{V}$, provided that the common coupling weight
$w$ is large enough.


\subsection{The position and velocity neighbor graph is unconnected}

In the main results, we assume that the position and velocity
neighbor graph
$\mathcal{G}\,=\,(\mathcal{V},\mathcal{E},\textbf{W})$ is
connected. In reality, however, it is not always the case. Suppose
that the graph consists of several connected subgraphs
$\mathcal{G}^{j}\,=\,(\mathcal{V}^{j},\mathcal{E}^{j},\textbf{W}^{j})\;(j\,\in\,\{1,2,\cdots\})$,
where $\bigcup\limits_j\,\mathcal{V}^{j}\,=\,\mathcal{V}$,
$\bigcup\limits_j\,\mathcal{E}^{j}\,=\,\mathcal{E}$,
$\textbf{W}^{j}\,\in\,R^{N_j\,\times\,N_j}$ is the weight matrix
of subgraph $\mathcal{G}^{j}$ which is symmetric, irreducible and
diffusive, and $\sum\limits_j\,N_j\,=\,N$. In subgraph
$\mathcal{G}^{j}$, pseudo-leader set $\bar{\mathcal{V}}^j$ can be
picked out according to Theorem 1 or Theorem 2. Put the
pseudo-leaders in all the subgraphs together, the scheme for
choosing pseudo-leader set
$\bar{\mathcal{V}}\,=\,\bigcup\limits_j\,\bar{\mathcal{V}}^{j}$ of
the whole unconnected group is obtained.

As discussed in the previous subsection, one single pseudo-leader
in each connected subgroup is enough for flocking. If, however,
there exists a connected subgroup in which no agent is informed,
it is impossible for this subgroup to detect the desired moving
information. Thus flocking failure, such as regular fragmentation,
may occur under this circumstance.


\subsection{The position and velocity neighbor graph is switching}

The aforementioned results are on the basis of a multi-agent group
whose topology is fixed. We can also consider a switching position
and velocity neighbor graph
$\mathcal{G}(t)\,=\,(\mathcal{V},\mathcal{E}(t),\textbf{W}(t))$,
where

$$
\textbf{W}(t)\,=\,\textbf{W}(\tau_k)\ \ \ \ \ \ \ \ \ \ \ \ \
t\,\in\,[\tau_k,\tau_{k+1}),\;k=0,1,\cdots
$$

\noindent with $\tau_0\,=\,0$ is a switching matrix. It is easy to
see that the graph switches at some instant time. Since
$\textbf{W}(\tau_k)$ is invariant in $[\tau_k,\tau_{k+1})$, a
temporal pseudo-leader set $\bar{\mathcal{V}}(\tau_k)$ in this
time interval can be determined based on Theorem 1 or Theorem 2.
Thus for a multi-agent group whose topology is switching, a
switching scheme can be employed to choose pseudo-leaders for
flocking.

For those switching models which utilize local information
(nearest neighbor law) to establish the group topology
\cite{OS2006}-\cite{HSS_2}, this technique will be more efficient
than using a fixed pseudo-leader set.


\subsection{Center of the group members}

Define the position and velocity of the center of all group
members as

$$
\begin{array}{l}
\bar{\textbf{p}}\,\triangleq\,\frac{1}{N}\,\sum\limits_{i=1}^N\,\textbf{p}_i,\\
\bar{\textbf{v}}\,\triangleq\,\frac{1}{N}\,\sum\limits_{i=1}^N\,\textbf{v}_i.
\end{array}
$$

\noindent Then we have the error system

$$
\begin{array}{l}
\bar{\textbf{e}}^\textbf{p}\,\triangleq\,\bar{\textbf{p}}\,-\,\textbf{p}_l\,
=\,\frac{1}{N}\,\sum\limits_{i=1}^N\,\textbf{e}_i^{\textbf{p}},\\
\bar{\textbf{e}}^\textbf{v}\,\triangleq\,\bar{\textbf{v}}\,-\,\textbf{v}_l\,
=\,\frac{1}{N}\,\sum\limits_{i=1}^N\,\textbf{e}_i^{\textbf{v}}.
\end{array}
$$

\noindent According to the proof of Theorem 1, every solution of the
system converges to the largest invariant set $\Omega^*$ of the set
$\{(\textbf{p}_i-\textbf{p}_j,\textbf{e}_i^\textbf{v})\,|\,\dot{L}=0\}$,
in which
$\dot{\textbf{e}}_i^\textbf{p}\,=\,\textbf{e}_i^\textbf{v}\,=\,\textbf{0}$
for $1\,\leq\,\,i\,\leq\,N$. This results in
$\dot{\bar{\textbf{e}}}^\textbf{p}\,=\,\bar{\textbf{e}}^\textbf{v}\,=\,\textbf{0}$
in $\Omega^*$. That is, in steady state, the center of all agent
velocities is equal to the desired velocity, and the position
difference between the center and the virtual leader remains
unchanged.


\section{Numerical Simulations}

\subsection{Flocking in a small-sized multi-agent group}

In the following, Theorem 1 is illustrated by using L\"{u} system
\cite{Lv2002} as the dynamical acceleration in system (1) and (2).
As a typical benchmark chaotic system, L\"u system is given by

$$
\begin{array}{rcl}
  \dot{\textbf{x}}
  & = & \left(\begin{array}{ccc}
              -a &  a  &  0 \\
               0 &  c  &  0 \\
               0 &  0  & -b
               \end{array}\right)
               \left(\begin{array}{c}
               x_{1} \\
               x_{2} \\
               x_{3}
               \end{array}\right)\,+\,
               \left(\begin{array}{c}
               0 \\
               -x_{1}x_{3} \\
               x_{1}x_{2}
               \end{array}\right) \\
  & \triangleq & {\textbf{R}}{\mathbf{x}}\,+\,\textbf{T}(\textbf{x})\,,
\end{array}
$$

\noindent which has a chaotic attractor when $a=36$, $b=3$, $c=20$.
For any two state vectors ${\textbf{y}}$ and ${\textbf{z}}$ of L\"u
system, there exist constants $M_s$ such that
$\|y_{s}\|,\,\|z_s\|\,\leq\,M_s$ for $1\,\leq\,s\,\leq\,3$ since the
L\"u attractor is bounded within a certain region. From simple
numerical calculation, $M_1=25$, $M_2=30$, $M_3=45$ is obtained.
Therefore, one has

$$
\begin{array}{l}
  (\textbf{y}-\textbf{z})^{\top}(\textbf{f}(\textbf{y})-\textbf{f}(\textbf{z}))\\
  =
  (\textbf{y}-\textbf{z})^{\top}\,\textbf{R}\,(\textbf{y}-\textbf{z})
  +\,(\textbf{y}-\textbf{z})^{\top}\,(\textbf{T}(\textbf{y})-\textbf{T}(\textbf{z}))\\
  =
  (\textbf{y}-\textbf{z})^{\top}\,\textbf{R}\,(\textbf{y}-\textbf{z})
  +\,(y_1\,-\,z_1)\,(z_2\,y_3\,-\,y_2\,z_3)\,\\
  =
  (\textbf{y}-\textbf{z})^{\top}\,
               \left(
               \left(\begin{array}{ccc}
               -a &  \frac{a}{2}  &  0 \\
               \frac{a}{2} &  c  &  0 \\
               0 &  0  & -b
               \end{array}\right)\,+\,
               \left(\begin{array}{ccc}
               0 &  -\frac{y_3}{2}  &  \frac{y_2}{2} \\
               -\frac{y_3}{2} &  0  &  0 \\
               \frac{y_2}{2} &  0  & 0
               \end{array}\right)\,\right)
               \,(\textbf{y}-\textbf{z})\\
  \leq
  \lambda_{max}\left(
  \begin{array}{ccc}
  -a &  \frac{a}{2}-\frac{y_3}{2}  &  \frac{y_2}{2} \\
  \frac{a}{2}-\frac{y_3}{2} &  c  &  0 \\
  \frac{y_2}{2} &  0  & -b
  \end{array}\right)\,
  \|(\textbf{y}-\textbf{z})\|^2\\
  \approx
  60.3402\,\|(\textbf{y}-\textbf{z})\|^2\,
  \end{array}
$$

Thus L\"u system satisfies the assumption A1 with
$\alpha\,=\,60.3402$.

For simplicity, consider the group with ten nodes whose adjacent
matrix is

$$
\textbf{B}\,=\,\left(\begin{array}{cccccccccc}
    -7  &   1  &   1  &   0  &   1  &   0  &   1  &   1  &   1  &   1\\
     1  &  -4  &   0  &   0  &   0  &   1  &   1  &   1  &   0  &   0\\
     1  &   0  &  -4  &   1  &   0  &   0  &   0  &   0  &   1  &   1\\
     0  &   0  &   1  &  -3  &   1  &   1  &   0  &   0  &   0  &   0\\
     1  &   0  &   0  &   1  &  -2  &   0  &   0  &   0  &   0  &   0\\
     0  &   1  &   0  &   1  &   0  &  -2  &   0  &   0  &   0  &   0\\
     1  &   1  &   0  &   0  &   0  &   0  &  -2  &   0  &   0  &   0\\
     1  &   1  &   0  &   0  &   0  &   0  &   0  &  -2  &   0  &   0\\
     1  &   0  &   1  &   0  &   0  &   0  &   0  &   0  &  -2  &   0\\
     1  &   0  &   1  &   0  &   0  &   0  &   0  &   0  &   0  &  -2
\end{array}\right).
$$


\noindent The common weight coupling is $w=70$. By deleting the
first three row-column pairs of $\textbf{B}$, the maximum
eigenvalue of the minor matrix is
$\lambda_{max}(\textbf{B}_7)\,=\,-1$. It is easy to see that

$$
\lambda_{max}(\textbf{B}_7)\,<\,-\frac{\alpha}{w}.
$$

\noindent Thus the first three agents can be picked out as
pseudo-leaders in the group according to Theorem 1.

Simulation results are shown in Fig. 1 - Fig. 4 . In the simulation,
the initial positions of the ten agents in the group are distributed
randomly from the cube $[0,5]^3$. The initial velocity coordinates
are randomly chosen from the cube $[0,2]^3$. The initial position
and velocity of the virtual leader, which is marked with a red star
in Fig. 1 and Fig. 2, are set as $\textbf{p}_l(0)\,=\,(6,6,6)^\top$
and $\textbf{v}_l(0)\,=\,(1,1,1)^\top$ respectively. Other
parameters are chosen as $h_{i_r}(0)\,=\,1$ and
$k_{i_r}\,=\,0.01\;(1\,\leq\,r\,\leq\,m,\;1\,\leq\,i_r\,\leq\,N)$.
Letting $\sigma\,=\,0.1$ in $\sigma$-norm and
$c_1\,=\,c_2\,=\,\frac{1}{2}$, we use the following APF

$$
V_{ij}(\|\textbf{p}_i-\textbf{p}_j\|_\sigma)\,=\,c_1\,\ln\,\|\textbf{p}_i-\textbf{p}_j\|_\sigma^2\,
+\,\frac{c_2}{\|\textbf{p}_i-\textbf{p}_j\|_\sigma^2}.
$$

The group's moving states at $t=0$ and $t=30$ are illustrated in
Fig. 1 and Fig. 2. Here, the solid circles represent the
pseudo-leaders that receive moving information of the virtual
leader, and the hollow ones denote the followers in the group. The
arrows display velocity vectors of all the agents. The dash lines
depict the connections between agents. From these two figures, it is
seen that in spite of the initial disordered state, all the agents
flock at $t=30$.

\begin{figure}
\begin{center}
\includegraphics[height=0.32\textwidth]{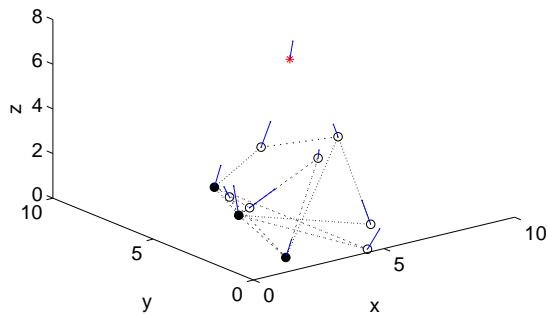}
\end{center}
\caption {Initial moving state of the group when $N=10$ and $m=3$.}
\end{figure}

\begin{figure}
\begin{center}
\includegraphics[height=0.32\textwidth]{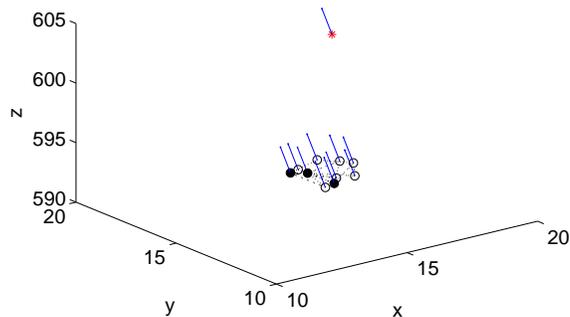}
\end{center}
\caption {Group's moving state when $N=10$ and $m=3$ at $t=30$.}
\end{figure}

The state of position and velocity differences between the agents
and the virtual leader are illustrated in Fig. 3 and Fig. 4
respectively. It is shown that the velocities of all the group
members converge to the desired velocity. Moreover, the position
errors between agents remain fixed after a period of time.

\begin{figure}
\begin{center}
\includegraphics[height=0.35\textwidth]{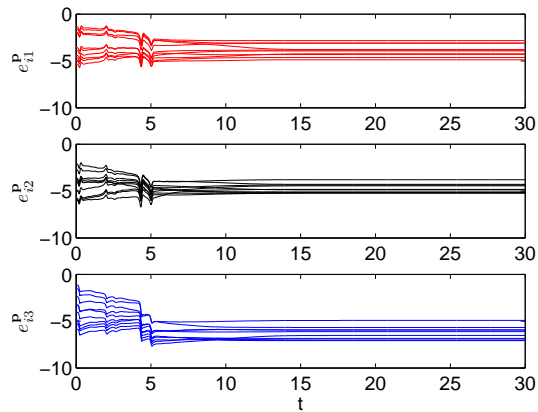}
\end{center}
\caption {The position differences ${e}_{is}^\textbf{p}$
$(1\,\leq\,i\,\leq\,10,\;1\,\leq\,s\,\leq\,3)$ between the agents
and the virtual leader when $N=10$ and $m=3$.}
\end{figure}

\begin{figure}
\begin{center}
\includegraphics[height=0.35\textwidth]{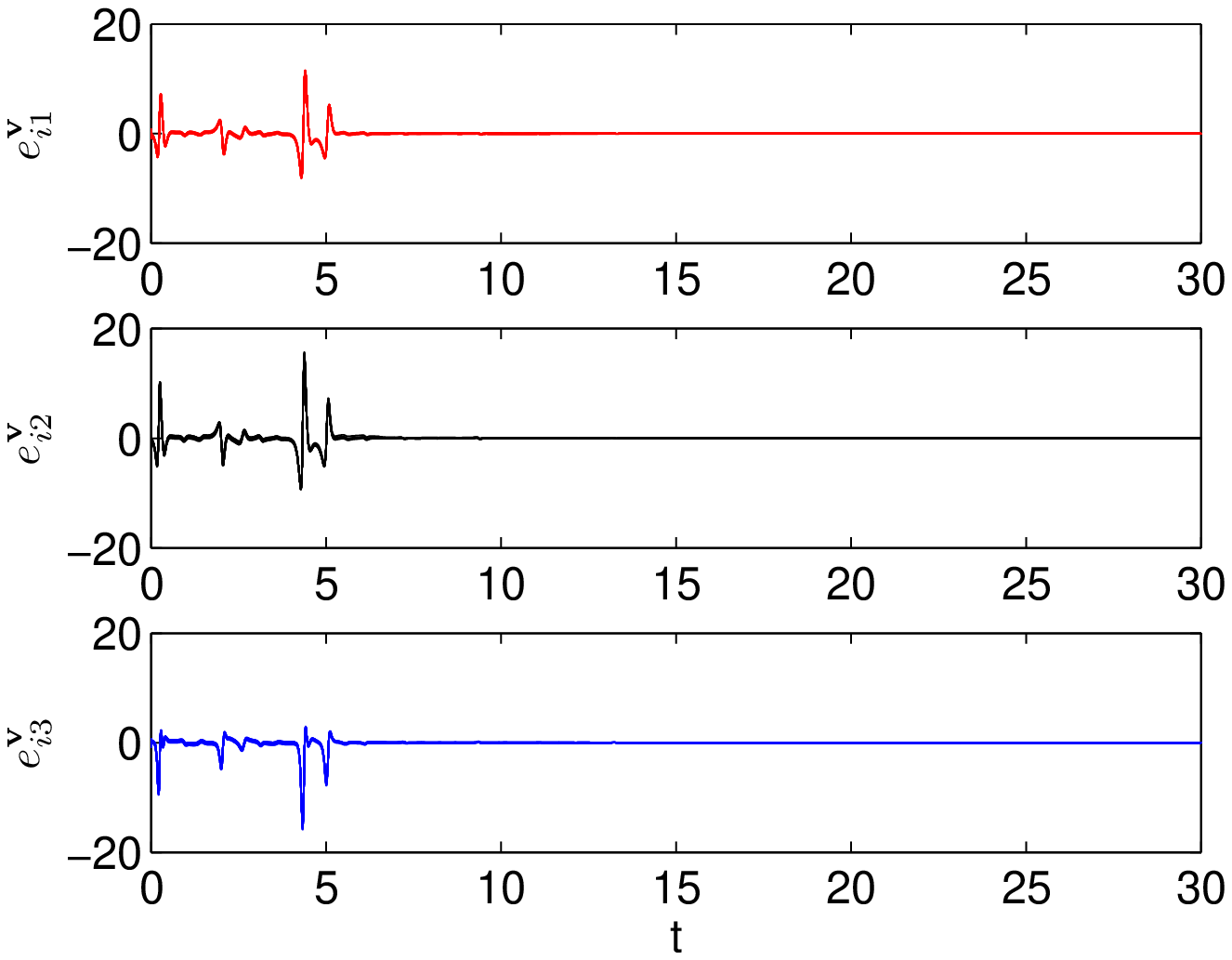}
\end{center}
\caption {The velocity differences ${e}_{is}^\textbf{v}$
$(1\,\leq\,i\,\leq\,10,\;1\,\leq\,s\,\leq\,3)$ between the agents
and the virtual leader when $N=10$ and $m=3$.}
\end{figure}


\subsection{Flocking in a 200-sized multi-agent group}

Large-scale groups with complex topology are common in nature. It
has been demonstrated that the topological information of most
large-sized systems display scale-free features, among which
Barab\'{a}si and Albert (BA) model \cite{BA} of preferential
attachment has become the standard mechanism to explain the
emergence of scale-free networks. Nodes are added to the network
with a preferential bias toward attachment to nodes with already
high degree. This naturally gives rise to hubs with a degree
distribution following a power-law.

In this subsection, a BA scale-free network consisting of $200$
agents with $q_0=5$ and $q=5$ are considered, where $q_0$ is the
size of the initial network, and $q$ is the number of edges added
in each step. Similar to the previous simulation, we take the
L\"{u} system as the dynamical acceleration in system (1) and (2).
The initial positions and the velocities of the $200$ agents are
selected randomly from the cube $[0,50]^3$ and $[0,2]^3$, and that
of the virtual leader are selected as
$\textbf{p}_l(0)\,=\,(60,60,60)^\top$ and
$\textbf{v}_l(0)\,=\,(1,1,1)^\top$. Assume that the other
parameters in the simulation are the same as those in the previous
subsection.

Denote $\textbf{B}^{BA}_{194}$ as the minor matrix of the adjacent
matrix $\textbf{B}^{BA}$ by removing $6$ row-column pairs which
corresponds to the $6$ agents with the largest degree in the whole
group. Since the maximum eigenvalue of $\textbf{B}^{BA}_{194}$
satisfies

$$
\lambda_{max}(\textbf{B}^{BA}_{194})\,=\,-0.9284\,<\,-0.8620\,=\,-\frac{\alpha}{w},
$$

\noindent the $6$ agents with the largest degrees can be picked out
as pseudo-leaders of the group. After using the control mechanism
presented in Theorem 1, flocking appears by letting just $3\%$
agents be informed. The moving states of all the group members at
$t=0$ and $t=30$ are exhibited in Fig. 5 and Fig. 6.

Clearly, the approaches presented in this paper are of high accuracy
with good performance not only for small-sized multi-agent groups
but also for larger-scale multi-agent systems.

\begin{figure}
\begin{center}
\includegraphics[height=0.32\textwidth]{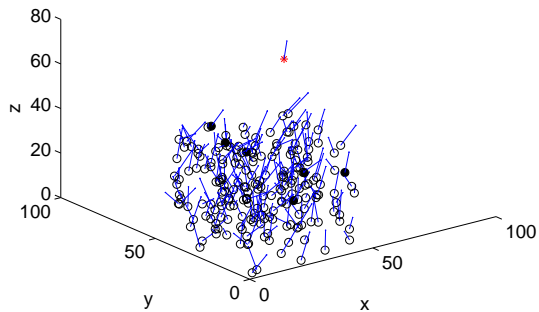}
\end{center}
\caption {Initial moving state of the group when $N=200$ and $m=6$.}
\end{figure}

\begin{figure}
\begin{center}
\includegraphics[height=0.32\textwidth]{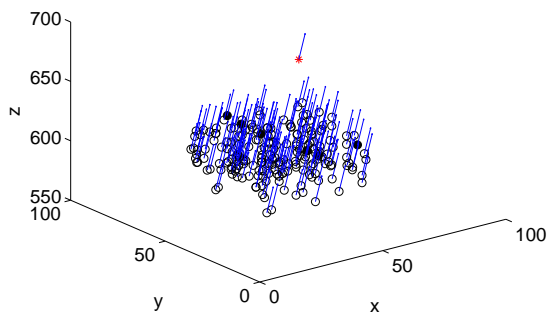}
\end{center}
\caption {Group's moving state when $N=200$ and $m=6$ at $t=30$.}
\end{figure}

\section{Conclusions}

In this paper, we have presented a criterion for choosing
pseudo-leaders in a multi-agent dynamical group. Particularly, the
weight configuration of the position and velocity neighbor graph is
not necessarily irreducible or time invariant. By combining the
ideas of virtual force and pseudo-leader mechanism, mathematical
analysis has been deduced to illustrate how to determine the
pseudo-leader set in a group. The proposed schemes have been proved
rigorously by using Schur complement and Lyapunov stability theory.
Finally, two computational examples including small-sized and
larger-scale multi-agent groups have been shown to illustrate the
effectiveness of the proposed approach.



\begin{thebibliography}{20}


{

\bibitem{Reynolds} C. W. Reynolds, ``Flocks, herds, and schools: a distributed behavioral model,"
{\it Computer Graphics,} vol. 21, no. 4, pp. 25-34, Jul. 1987.

\bibitem{Vicsek} T. Vicsek, A. Czirok, E. Ben-Jacob, I. Cohen and O. Shochet, ``Novel type of phase transition in
a system of self-driven particles," {\it Phys. Rev. Lett.,} vol.
75, pp. 1226-1229, Aug. 1995.

\bibitem{Shen2008} Jackie (Jianhong) Shen, ``Cucker-Smale flocking under hierarchical leadership,"
{\it SIAM Journal on Applied Mathematics}, vol. 68, no. 3, pp.
694-719, 2007.

\bibitem{Tanner2003_1} H. G. Tanner, A. Jadbabaie and G. J. Pappas, ``Stable flocking of mobile agents, part I:
fixed topology," {\it Proc. the 42nd IEEE Conference on Decision
and Control,} pp. 2010-2015, Dec. 2003.

\bibitem{Tanner2003_2} H. G. Tanner, A. Jadbabaie and G. J. Pappas, ``Stable flocking of mobile agents, part II:
dynamic topology," {\it Proc. the 42nd IEEE Conference on Decision
and Control,} pp. 2010-2015, Dec. 2003.

\bibitem{Tanner2007} H. G. Tanner, A. Jadbabaie and G. J. Pappas, ``Flocking in fixed and switching networks,"
{\it IEEE Transactions on Automatic Control,} vol. 52, no. 5, pp.
863-868, May. 2007.

\bibitem{OS2006} R. Olfati-Saber, ``Flocking for multi-agent dynamic systems: algorithms and theory,"
{\it IEEE Transactions on Automatic Control,} vol. 51, no. 3, pp.
401-420, Mar. 2006.

\bibitem{HSS_1} H. S. Su, X. F. Wang and Z. L. Lin, ``Flocking of multi-agents with a
virtual leader, part I: with a minority of informed agents," {\it
Proc. the 46th IEEE Conference on Decision and Control,} pp.
2937-2942, Dec. 2007.

\bibitem{HSS_2} H. S. Su, X. F. Wang and Z. L. Lin, ``Flocking of multi-agents with a
virtual leader, part II: with a virtual leader of varying
velocity," {\it Proc. the 46th IEEE Conference on Decision and
Control,} pp. 1429-1434, Dec. 2007.

\bibitem{HS} H. Shi, L. Wang and T. G. Chu, ``Virtual leader approach to
coordinated control of multiple mobile agents with asymmetric
interactions," {\it Physica D,} vol. 213, pp. 51-65, 2006.

\bibitem{XHLi} X. H. Li, J. Z. Xiao and Z. J. Cai, ``Stable flocking of swarms using
local information," {\it Proc. IEEE Int. Conf. on Systems, Man and
Cybernetics,} pp. 3921-3926, Oct. 2005.




\bibitem{Godsil} C. Godsil and G. Royle, {\it Algebraic Graph
Theory.} Springer-Verlag, New York, 2001.

\bibitem{Horn} P. A. Horn and C. R. Johnson, {\it Matrix Analysis.} Cambridge University Press, New York, 1985.

\bibitem{Gradshteyn} I. S. Gradshteyn and I. M. Ryzhik, {\it Tables of Integrals, Series, and Products.}
6th edition, Academic Press, 2000.

\bibitem{Hassan2002} K. K. Hassan, {\it Nonlinear systems.} 3rd edition, Prentice Hall,
2002.

\bibitem{Boyd} S. Boyd, L. E. Ghaoui, E. Feron and V.
Balakrishnan, {\it Linear matrix inequalities in system and
control theory.} Philadelphia, PA: SIAM, 1994.

\bibitem{WWYu2008} W. W. Yu, J. D. Cao and J. H. L\"{u}, ``Global synchronization of linearly hybrid
coupled networks with time-varying delay," {\it SIAM Journal on
Applied Dynamical Systems}, vol. 7, no. 1, pp. 108-133, 2008.

\bibitem{WLLu} T. P. Chen, X. W. Liu and W. L. Lu, ``Pinning complex networks by a
single controller," {\it IEEE Trans. Circuits Syst. I}, vol. 54, no.
6, pp. 1317-1326, 2007.

\bibitem{Jeffreys} H. Jeffreys and B. S. Jeffreys, {\it Methods
of Mathematical Physics.} 3rd edition, Cambridge University Press,
1988.

\bibitem{Lv2002} J. H. L\"u, G. Chen, D. Cheng, and S. Celikovsky, ``Bridge
the gap between the Lorenz system and the Chen system," {\it
International Journal of Bifurcation and Chaos}, vol. 12, no. 12,
pp. 2917-2926, Dec. 2002.

\bibitem{BA} A-L. Barab$\acute{a}$si and R. Albert, ``Emergence of Scaling in Random Networks," {\it Science},
vol. 286, no. 5439, pp. 509-512, Oct. 1999.






}


\end{thebibliography}
\end{document}